\documentclass[12pt, fleqn, a4paper]{article} 
\usepackage{hyperref}
\hypersetup{pdfstartview = FitH,
            pdfborder    = {0 0 0.5},
            colorlinks   = false
}
 \usepackage{epsfig, subcaption}
 \usepackage{multirow}
 \usepackage{clshan-math, clshan-dm-dd, clshan-dm-ddd}
 \def \lsim {\:\raisebox{-0.7 ex}{$\stackrel{\textstyle<}{\sim}$}\:}
 \def \gsim {\:\raisebox{-0.7 ex}{$\stackrel{\textstyle>}{\sim}$}\:}
 \def \PlotNumberAa {00000}
 \def \PeriodCa     {19.49  --  79.49}
 \def \PeriodCb     {110.74 -- 170.74}
 \def \PeriodCc     {201.99 -- 261.99}
 \def \PeriodCd     {293.24 -- 353.24}
 \def \PlotNumberCa {04949}
 \def \PlotNumberCb {14074}
 \def \PlotNumberCc {23199}
 \def \PlotNumberCd {32324}
\newcommand{\rmF}   {\rmXA{F}  {19}}
\newcommand{\rmAr}  {\rmXA{Ar} {40}}
\newcommand{\rmGe}  {\rmXA{Ge} {73}}
\newcommand{\rmXe}  {\rmXA{Xe}{129}}
\newcommand{\rmW}   {\rmXA{W} {183}}
\newcommand{\InsertSKPPlotS} [3] [10.5] {
\begin{figure} [t!]
\begin{center}
 \includegraphics [width = #1 cm] {skp-#2}
\end{center}
\caption{
 #3
}
\label{fig:#2}
\end{figure}
}
\newcommand{\OnlinePlotfv} [8] {
\href{http://www.tir.tw/phys/hep/dm/amidas-2d/amidas-2d.php%
      ?amidas_2D_function=fv_eff%
      &mode_fv_eff=#3%
      &frame=#4%
      &mode_animation=#5%
      &period=#6%
      &event_No=#7}
     {#8}%
}
\newcommand{\Rowfv} [3] {
\begin{subfigure} [c] {17.19 cm}
\begin{center}
  \OnlinePlotfv
   {}
   {} {N_v}
   {\ShortFrame}
   {#3&target=#1}
   {periodA}
   {500}
   {\includegraphics [width = 4.1 cm]
     {N_v-#1-\WIMPmass-\ShortFrame-\EventNumber-\PlotNumber}}%
  \OnlinePlotfv
   {}
   {} {N_ang}
   {\ShortFrame}
   {#3&target=#1}
   {periodA}
   {500}
   {\includegraphics [width = 4.1 cm]
     {N_ang-#1-\WIMPmass-\ShortFrame-\EventNumber-\PlotNumber}}%
  \OnlinePlotfv
   {}
   {} {P_ang}
   {\ShortFrame}
   {#3&target=#1}
   {periodA}
   {500}
   {\includegraphics [width = 4.1 cm]
     {P_ang-#1-\WIMPmass-\ShortFrame-\EventNumber-\PlotNumber}}%
  \OnlinePlotfv
   {}
   {} {PoN_ang}
   {\ShortFrame}
   {#3&target=#1}
   {periodA}
   {500}
   {\includegraphics [width = 4.1 cm]
     {PoN_ang-#1-\WIMPmass-\ShortFrame-\EventNumber-\PlotNumber}}%
\vspace{-0.4 cm}
\end{center}
\caption{#2}
\end{subfigure}
}
\newcommand{\InsertPlotfv} [2] {
\begin{figure} [t!]
\begin{center}
 \Rowfv {F19}   {$\rmF$}  {#1} \\ \vspace{ 0.2 cm}
 \Rowfv {Ar40}  {$\rmAr$} {#1} \\ \vspace{ 0.2 cm}
 \Rowfv {Ge73}  {$\rmGe$} {#1} \\ \vspace{ 0.2 cm}
 \Rowfv {Xe129} {$\rmXe$} {#1} \\ \vspace{ 0.2 cm}
 \Rowfv {W183}  {$\rmW$}  {#1} \\ \vspace{-0.4 cm}
\end{center}
\caption{
 #2
}
\label{fig:fv-\WIMPmass-\ShortFrame-\EventNumber-\PlotNumber}
\end{figure}
}
\newcommand{\RowfvAnnual} [3] {
\begin{subfigure} [c] {17.19 cm}
\begin{center}
  \OnlinePlotfv
   {}
   {} {N_v}
   {\ShortFrame}
   {annual&target=\Target&mchi=#1}
   {periodC}
   {500}
   {\includegraphics [width = 4.1 cm]
     {N_v-\Target-\WIMPmass-\ShortFrame-\EventNumber-#2}}%
  \OnlinePlotfv
   {}
   {} {N_ang}
   {\ShortFrame}
   {annual&target=\Target&mchi=#1}
   {periodC}
   {500}
   {\includegraphics [width = 4.1 cm]
     {N_ang-\Target-\WIMPmass-\ShortFrame-\EventNumber-#2}}%
  \OnlinePlotfv
   {}
   {} {P_ang}
   {\ShortFrame}
   {annual&target=\Target&mchi=#1}
   {periodC}
   {500}
   {\includegraphics [width = 4.1 cm]
     {P_ang-\Target-\WIMPmass-\ShortFrame-\EventNumber-#2}}%
  \OnlinePlotfv
   {}
   {} {PoN_ang}
   {\ShortFrame}
   {annual&target=\Target&mchi=#1}
   {periodC}
   {500}
   {\includegraphics [width = 4.1 cm]
     {PoN_ang-\Target-\WIMPmass-\ShortFrame-\EventNumber-#2}}%
\vspace{-0.4 cm}
\end{center}
\caption{#3}
\end{subfigure}
}
\newcommand{\InsertPlotfvAnnual} [2] {
\begin{figure} [t!]
\begin{center}
 \RowfvAnnual {#1} {\PlotNumbera} {\Perioda\ day} \\ \vspace{ 0.2 cm}
 \RowfvAnnual {#1} {\PlotNumberb} {\Periodb\ day} \\ \vspace{ 0.2 cm}
 \RowfvAnnual {#1} {\PlotNumberc} {\Periodc\ day} \\ \vspace{ 0.2 cm}
 \RowfvAnnual {#1} {\PlotNumberd} {\Periodd\ day} \\ \vspace{-0.4 cm}
\end{center}
\caption{
 #2
}
\label{fig:fv-\Target-\WIMPmass-\ShortFrame-\EventNumber-\PlotNumbera}
\end{figure}
}
\begin{document}
\thispagestyle{empty}
\begin{flushright}
 March 2021
\end{flushright}
\begin{center}
{\Large\bf
 3-Dimensional Effective Velocity Distribution of       \\
 Halo Weakly Interacting Massive Particles              \\ \vspace{0.2 cm}
 Scattering off Nuclei in Direct Dark Matter Detectors} \\
\vspace*{0.7 cm}
 {\sc Chung-Lin Shan}                                   \\
\vspace{0.5 cm}
 {\small\it
  Preparatory Office of
  the Supporting Center for
  Taiwan Independent Researchers                        \\ \vspace{0.05 cm}
  P.O.BOX 21 National Yang Ming Chiao Tung University,
  Hsinchu City 30099, Taiwan, R.O.C.}                   \\~\\~\\
 {\it E-mail:} {\tt clshan@tir.tw}
\end{center}
\vspace{2 cm}
\begin{abstract}

 In this paper,
 as the third part of the third step of
 our study on developing data analysis procedures
 for using 3-dimensional information
 offered by directional direct Dark Matter detection experiments
 in the future,
 we introduce
 a 3-dimensional
 {\em effective} velocity distribution of
 halo Weakly Interacting Massive Particles (WIMPs),
 which,
 instead of the theoretically prediction of
 the entire Galactic Dark Matter particles,
 describes the actual
 velocity distribution of WIMPs
 scattering off (specified) target nuclei
 in an underground detector.
 Its target and WIMP--mass dependences
 as well as
 (``annual'' modulations of)
 its ``anisotropy''
 in the Equatorial/laboratory
 and even the Galactic coordinate systems
 will be demonstrated and discussed in detail.
 For readers' reference,
 all simulation plots presented in this paper (and more)
 can be found ``in animation''
 on our online (interactive) demonstration webpage
 ({\tt \url{http://www.tir.tw/phys/hep/dm/amidas-2d/}}).

\end{abstract}
\clearpage
\section{Introduction}

 In the last (more than)
 three decades,
 a large number of experiments has been built
 and is being planned
 to search for the most favorite Dark Matter (DM) candidate:
 Weakly Interacting Massive Particles (WIMPs) $\chi$,
 by
 direct detection of
 the scattering recoil energy
 of ambient WIMPs off target nuclei
 in low--background underground laboratory detectors
 (see Refs.~%
  \cite{SUSYDM96,
        Gaitskell04,
        Baudis12c,
        Baudis20}
  for reviews).

 Besides non--directional direct detection experiments
 measuring only recoil energies
 deposited in detectors,
 the ``directional'' detection of Galactic DM particles
 has been proposed more than one decade
 to be a promising experimental strategy
 for discriminating signals from backgrounds
 by using additional 3-dimensional information
 (recoil tracks and/or head--tail senses)
 of (elastic) WIMP--nucleus scattering events
 (see Refs.~%
  \cite{Ahlen09,
        Mayet16, Battat16b,
        Vahsen20,
        Vahsen21}
  for reviews and recent progresses).

\begin{figure} [t!]
\begin{center}
 \includegraphics [width = 15.5 cm] {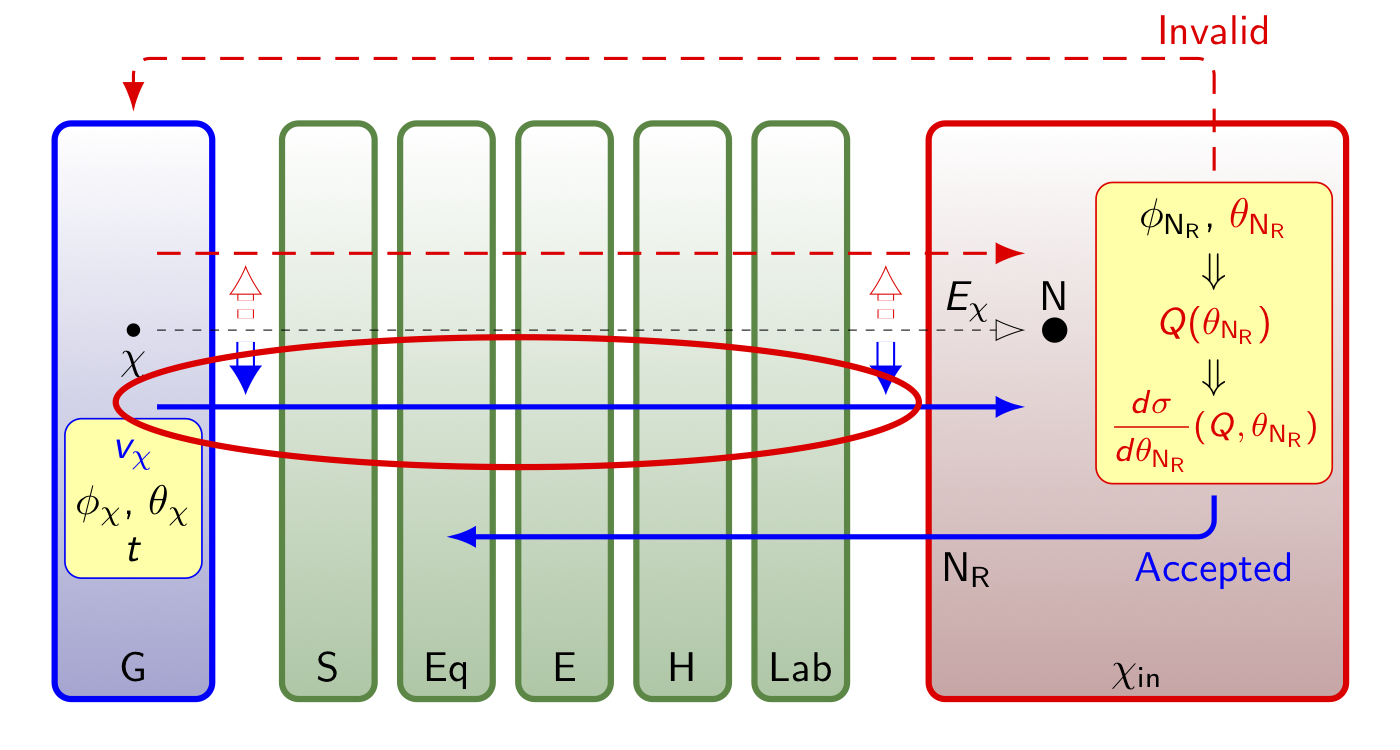}
\end{center}
\caption{
 The workflow of
 our double--Monte Carlo simulation and data analysis procedure of
 3-dimensional elastic WIMP--nucleus scattering.
 See the text for detailed descriptions.
}
\label{fig:workflow}
\end{figure}

 As the preparation
 for our future study
 on the development of data analysis procedures
 for using and/or combining 3-D information
 offered by directional Dark Matter detection experiments
 to,
 e.g.,
 reconstruct the 3-dimensional WIMP velocity distribution,
 in Ref.~\cite{DMDDD-N},
 we started with the Monte Carlo generation of
 the 3-D velocity of
 (incident) halo WIMPs
 in the Galactic coordinate system,
 including
 the magnitude,
 the direction,
 and the incoming/scattering time.
 Each generated 3-D WIMP velocity
 has then been transformed to
 the laboratory--independent
 (Ecliptic,
  Equatorial,
  and Earth)
 coordinate systems
 as well as
 to the laboratory--dependent
 (horizontal and laboratory)
 coordinate systems
 (see the simulation workflow
  sketched in Fig.~\ref{fig:workflow})
 for the investigations on
 the angular distribution patterns of
 the 3-D WIMP velocity (flux)
 and the (accumulated and average) kinetic energy
 in different celestial coordinate systems
 \cite{DMDDD-N, DMDDD-P}
 as well as
 the Bayesian reconstruction of
 the radial component (magnitude) of
 the 3-D WIMP velocity
 \cite{DMDDD-N}.
 Not only
 the diurnal modulations,
 we demonstrated also
 the ``annual'' modulations of
 the angular WIMP velocity (flux)/kinetic energy distributions
 \cite{DMDDD-N, DMDDD-P}.

 However,
 besides recoil energies,
 what one could measure (directly)
 in directional DM detection experiments
 is recoil tracks (with the sense--recognition)
 and in turn recoil angles (directions)
 of scattered target nuclei.
 In Ref.~\cite{DMDDD-3D-WIMP-N},
 we have finally achieved
 our double--Monte Carlo scattering--by--scattering simulation of
 the 3-D elastic WIMP--nucleus scattering process
 and can provide
 the 3-D recoil direction and then the recoil energy of
 the WIMP--scattered target nuclei
 event by event
 in different celestial coordinate systems.
 Then,
 in Ref.~\cite{DMDDD-NR},
 we have demonstrated
 the simulation results of
 the angular distributions of
 the recoil direction (flux)
 as well as
 the accumulated and the average recoil energies of
 the target nuclei
 scattered by incident halo WIMPs
 (indicated by the lower solid blue arrow
  in Fig.~\ref{fig:workflow}).

 During the study on
 the angular distributions of
 the recoil direction (flux)/energy of
 target nuclei
 scattered by incident halo WIMPs
 \cite{DMDDD-NR},
 several questions have came to our mind:
 does the subgroup of WIMPs
 scattering off target nuclei
 (circled in the simulation workflow
  in Fig.~\ref{fig:workflow})
 have the same 3-dimensional velocity distribution
 as the main group of the entire halo WIMPs
 (impinging into a (directional) direct DM detector
  but not necessarily scattering off target nuclei)?
 Or,
 equivalently,
 does the WIMPs
 scattering off Ar or Xe nuclei
 have the same 3-D velocity distribution
 as the WIMPs
 scattering off Si or Ge nuclei?
 One could also ask that,
 once one can reconstruct
 the (3-D) velocity distribution of
 WIMPs
 by using (directional) direct detection data,
 is the reconstructed (3-D) velocity distribution
 indeed
 that of the entire halo WIMPs?

 So far
 (most) people would assume ``yes'' for
 these three ``simple'' questions
 as well as
 for (almost) all experimental data analyses
 (of the exclusion limits)
 in the parameter space of WIMP properties
 and
 for developing phenomenological methods
 to reconstruct WIMP properties
 (including our own earlier works).
 In this paper,
 as the counterpart of
 our study on
 the angular distributions of
 the recoil flux/energy of
 WIMP--scattered target nuclei
 \cite{DMDDD-NR},
 we study
 the 3-D velocity distribution of incident WIMPs
 ``scattering'' off target nuclei
 for finding out the answers of these questions.

 The remainder of this paper is organized as follows.
 In Sec.~2,
 we describe
 the overall workflow of
 our double--Monte Carlo
 scattering--by--scattering simulation procedure of
 3-dimensional elastic WIMP--nucleus scattering
 and review briefly
 the validation criterion of
 our MC simulation of
 3-D WIMP scattering events.
 Then,
 in Secs.~3 and 4,
 we present
 the (WIMP--mass and target--nucleus dependent)
 3-D (radial and angular) distributions of
 the WIMP {\em effective} velocity
 as well as
 the angular distributions of
 the corresponding (average) kinetic energy
 in the Equatorial and the Galactic coordinate systems,
 respectively.
 An {\em annual} modulation of
 (the {\em anisotropy} of)
 the 3-D WIMP effective velocity distribution
 in the {\em Galactic} coordinate system
 will especially be demonstrated and discussed in detail.
 We conclude in Sec.~5.

%
 %
%
\section{Monte Carlo scattering--by--scattering simulation of
         3-dimensional elastic WIMP--nucleus scattering events}
\label{sec:3D-WIMP-N}

 In Ref.~\cite{DMDDD-3D-WIMP-N},
 we have
\begin{enumerate}
\item
 reviewed
 the MC generation of
 the 3-D velocity information
 (the magnitude,
  the direction,
  and
  the incoming/scattering time)
 of Galactic WIMPs,
\item
 summarized
 the definitions of
 and
 the transformations between
 all celestial coordinate systems.
\end{enumerate}
 In this section,
 we describe at first
 the overall workflow of
 our Monte Carlo scattering--by--scattering simulation of
 the 3-D elastic WIMP--nucleus scattering process.
 Then
 we review briefly
 the validation criterion of
 our Monte Carlo simulations
 by taking into account
 the cross section (nuclear form factor) suppression
 on {\em each} generated recoil energy.

\subsection{Simulation workflow}
\label{sec:workflow}

 In this subsection,
 we describe
 the overall workflow of
 our double--Monte Carlo simulation and data analysis procedure of
 3-D elastic WIMP--nucleus scattering
 sketched in Fig.~\ref{fig:workflow} in detail:
\begin{enumerate}
\item
 The 3-D velocity information of incident halo WIMPs
 (the magnitude and the direction
  as well as
  the incoming/scattering time)
 is MC generated
 according to a specified model of the Dark Matter halo
 in the Galactic coordinate system
 (the blue subframe)
 \cite{DMDDD-N, DMDDD-3D-WIMP-N}.
\item
 The generated 3-D WIMP velocities
 will be transformed through
 the laboratory--independent
 (Ecliptic,
  Equatorial,
  and Earth)
 coordinate systems
 as well as
 the laboratory--dependent
 (horizontal and laboratory)
 coordinate systems
 (the green subframes)
 \cite{DMDDD-N, DMDDD-3D-WIMP-N}
 and at the end into the incoming--WIMP coordinate system
 (the red subframe)
 \cite{DMDDD-3D-WIMP-N}.
\item
 In the incoming--WIMP coordinate system,
 the 3-D elastic WIMP--nucleus scattering process
 will also be MC simulated
 by generating
 the orientation of the scattering plane $\phiNRchi$
 and the ``equivalent'' recoil angle $\thetaNRchi$
 (defined in Fig.~\ref{fig:NR-chi-Lab}).
 They define the recoil direction of
 the scattered target nucleus
 and the latter,
 combined with the transformed WIMP incident velocity,
 will then be used for estimating
 the transferred recoil energy to
 the target nucleus,
 $Q(\thetaNRchi)$,
 and the differential WIMP--nucleus scattering cross section
 with respect to
 the recoil angle,
 $d\sigma / d\thetaNRchi (Q, \thetaNRchi)$,
 in our event validation criterion
 (see Sec.~\ref{sec:dsigma_dthetaNRchi}).
\item
 The orientation of
 the scattering plane $\phiNRchi$
 and the equivalent recoil angle $\thetaNRchi$
 of the {\em accepted} recoil events
 will be transformed (back)
 through all considered celestial coordinate systems
 (indicated by the lower solid blue arrow).
 All these 3-D recoil information of
 the scattered target nucleus
 accompanied with
 the corresponding recoil energy $Q$
 as well as
 the 3-D velocity of the scattering WIMP
 in different coordinate systems
 (the upper solid blue arrow)
 will be recorded
 for further analyses%
\footnote{
 In this paper,
 we focus on investigating
 the 3-dimensional effective velocity distribution of
 the incident WIMPs
 scattering off target nuclei
 (indicated by the upper blue arrow).
 A detailed study on
 the angular distributions of
 the nuclear recoil direction (flux)/energy
 (the lower blue arrow)
 is presented
 in Ref.~\cite{DMDDD-NR}
 separately.
}.
\item
 For the {\em invalid} cases,
 in which
 the estimated recoil energies
 are out of the experimental measurable energy window
 or suppressed by the validation criterion,
 the generated 3-D information
 on the incident WIMP
 (the lower dashed red arrow)
 (and that on the scattered nucleus)
 will be discarded
 and the generation/validation process of
 one WIMP scattering event
 will be restarted from the Galactic coordinate system
 (the upper dashed red arrow).
\end{enumerate}
\subsection{Validation of 3-D elastic WIMP--nucleus scattering events}
\label{sec:dsigma_dthetaNRchi}
\InsertSKPPlotS
 {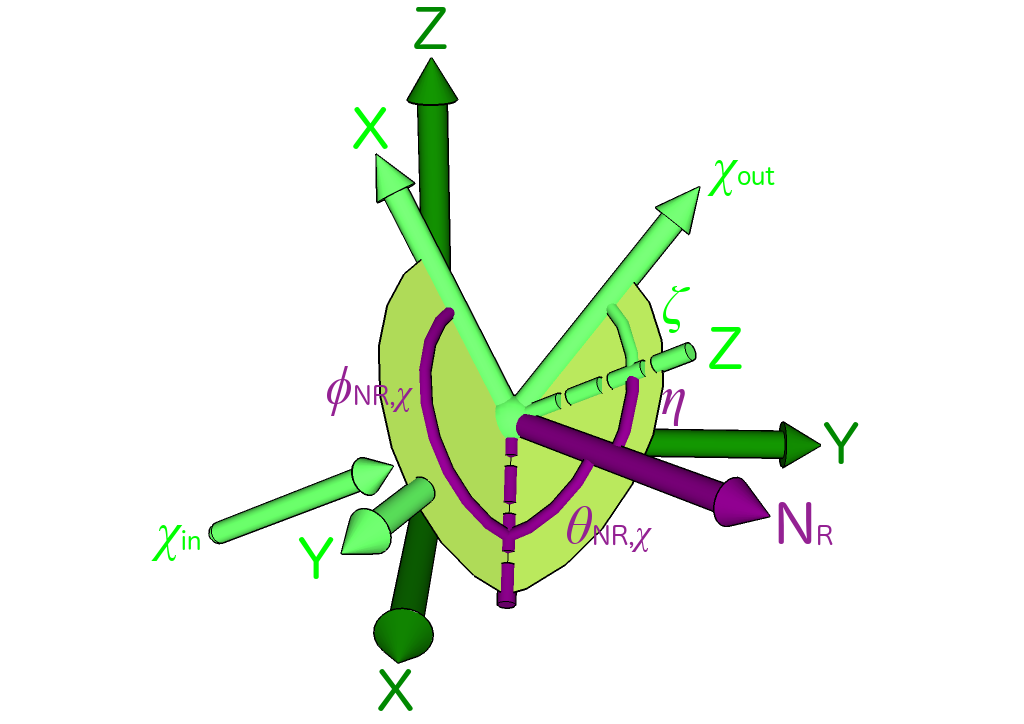}
 {A 3-D elastic WIMP--nucleus scattering event
  in the (light--green) incoming--WIMP
  and the (dark--green) laboratory coordinate systems
  \cite{DMDDD-N, DMDDD-3D-WIMP-N}.
  $\zeta$ and $\eta$ are
  the scattering angle of
  the outgoing WIMP $\chi_{\rm out}$
  and
  the recoil angle of
  the scattered target nucleus N$_{\rm R}$
  measured in the incoming--WIMP coordinate system
  of {\em this single} scattering event,
  respectively.
  While
  the azimuthal angle of
  the recoil direction of
  the scattered nucleus N$_{\rm R}$
  in this incoming--WIMP coordinate system,
  $\phiNRchi$,
  indicates the orientation of the scattering plane,
  the elevation of
  the recoil direction of N$_{\rm R}$,
  $\thetaNRchi$,
  is namely the complementary angle of
  the recoil angle
  $\eta$.%
  }

 In Fig.~\ref{fig:NR-chi-Lab},
 we sketch 
 the process of
 one single 3-D elastic WIMP--nucleus scattering event:
 $\chi_{\rm in/out}$ indicate
 the incoming and the outgoing WIMPs,
 respectively.
 While
 $\zeta$ indicates
 the scattering angle of
 the outgoing WIMP $\chi_{\rm out}$
 (measured from the $\zchi$--axis
  of the incoming--WIMP coordinate system,
  which is defined as
  the direction of the incident velocity of
  the incoming WIMP
  $\Vchi$),
 $\eta$ is the recoil angle of
 the scattered target nucleus N$_{\rm R}$.
 It can be found that
 the elevation of
 the recoil direction of
 the scattered nucleus,
 $\thetaNRchi$,
 is namely the complementary angle of
 the recoil angle $\eta$.
 Thus,
 in our simulations,
 we can use
\beq
      \thetaNRchi
 \in  [0,~\pi / 2]
\label{eqn:thetaNRchi_range}
\eeq
 as the ``equivalent'' recoil angle%
\footnote{
 Note that,
 without special remark,
 we will use hereafter simply ``the recoil angle''
 to indicate ``the equivalent recoil angle $\thetaNRchi$''
 (not $\eta$).
}.
 Since
 for one WIMP event
 transformed
 into the laboratory coordinate system
 with the velocity of
 $\Vchi$,
 the kinetic energy
 can be given by
\beq
     \Echi
  =  \frac{1}{2} \mchi |\Vchi|^2
  =  \frac{1}{2} \mchi \vchiLab^2
\~.
\label{eqn:Echi}
\eeq
 Then
 the recoil energy of the scattered target nucleus
 in the incoming--WIMP coordinate system
 can be estimated by
 the equivalent recoil angle
 $\thetaNRchi$
 as
 \cite{DMDDD-3D-WIMP-N}
\beq
     Q(\thetaNRchi)
  =  \bbrac{\afrac{2 \mrN^2}{\mN} \vchiLab^2}
     \sin^2(\thetaNRchi)
\~,
\label{eqn:QQ_thetaNRchi}
\eeq
 where
\(
         \mrN
 \equiv  \mchi \mN / (\mchi + \mN)
\)
 is the reduced mass of
 the WIMP mass $\mchi$ and
 that of the target nucleus $\mN$.
 And
 the differential cross section $d\sigma$
 given by
 the absolute value of the momentum transfer
 from the incident WIMP to the recoiling target nucleus,
 $q = |{\bf q}| = \sqrt{2 \mN Q}$,
 can be obtained as
 \cite{SUSYDM96, DMDDD-3D-WIMP-N}
\beq
     d\sigma
  =  \frac{1}{\vchiLab^2}
     \afrac{\sigma_0}{4 \mrN^2} F^2(q) \~ dq^2
  =  \sigma_0 F^2(Q(\thetaNRchi))
     \sin(2 \thetaNRchi) \~ d\thetaNRchi
\~.
\label{eqn:dsigma_thetaNRchi}
\eeq
 Hence,
 the differential WIMP--nucleus scattering cross section
 with respect to
 the recoil angle
 $\thetaNRchi$
 can generally be given by
 \cite{DMDDD-3D-WIMP-N}%
\footnote{
 It would be important to emphasize here that,
 to the best of our knowledge,
 this should be the first time in literature that
 some constraints on
 the nuclear recoil angle/direction
 caused by
 (elastic) WIMP--nucleus scattering cross sections (nuclear form factors)
 have been considered
 in (3-D) WIMP scattering simulations.
}
\beq
     \Dd{\sigma}{\thetaNRchi}
  =  \bbigg{  \sigmaSI F_{\rm SI}^2(Q(\thetaNRchi))
            + \sigmaSD F_{\rm SD}^2(Q(\thetaNRchi)) }
     \sin(2 \thetaNRchi)
\~.
\label{eqn:dsigma_dthetaNRchi}
\eeq
 Here
 $\sigma_0^{\rm (SI, SD)}$ are
 the spin--independent (SI)/spin--dependent (SD) total cross sections
 ignoring the form factor suppression
 and
 $F_{\rm (SI, SD)}(Q)$ indicate the elastic nuclear form factors
 corresponding to the SI/SD WIMP interactions,
 respectively.

 Finally,
 taking into account
 the proportionality of the WIMP flux
 to the incident velocity,
 the generating probability distribution of
 the recoil angle
 $\thetaNRchi$,
 which is proportional to
 the scattering event rate of
 incident halo WIMPs
 with an incoming velocity $\vchiLab$
 off target nuclei
 going into recoil angles of
 $\thetaNRchi \pm d\thetaNRchi / 2$
 with recoil energies of $Q \pm dQ / 2$,
 can generally be given by
 \cite{DMDDD-3D-WIMP-N}
\beq
     f_{{\rm N_R}, \chiin, \theta}(\thetaNRchi)
  =  \afrac{\vchiLab}{v_{\chi, {\rm cutoff}}}
     \bbigg{  \sigmaSI F_{\rm SI}^2(Q(\thetaNRchi))
            + \sigmaSD F_{\rm SD}^2(Q(\thetaNRchi)) }
     \sin(2 \thetaNRchi)
\~,
\label{eqn:f_NR_thetaNRchi}
\eeq
 where
 $v_{\chi, {\rm cutoff}} \simeq 800$ km/s is
 a cut--off velocity of incident halo WIMPs
 in the laboratory coordinate system.

%

%
 %
%
\section{3-D WIMP effective velocity distribution
         in the Equatorial frame}
\label{sec:fv_eff-Eq}

 In Refs.~\cite{DMDDD-N} and \cite{DMDDD-P},
 we demonstrated
 the angular distributions of
 the 3-D velocity (flux)
 and the kinetic energy of
 Galactic halo WIMPs
 impinging into (directional) direct DM detectors
 in different celestial coordinate systems,
 which show clearly
 the anisotropy and the directionality
 (the annual and the diurnal modulations).
 On the other hand,
 in Ref.~\cite{DMDDD-NR},
 we have presented
 (the anisotropy and the directionality (the annual modulation) of)
 the angular distributions of
 the recoil direction (flux)
 as well as
 the accumulated and the average recoil energies of
 several frequently used target nuclei
 scattered by (simulated) incident Galactic WIMPs
 observed in different coordinate systems.

 Then,
 in this and the next sections,
 we discuss correspondingly
 (the anisotropy and the directionality (the annual modulation) of)
 the 3-dimensional {\em effective} velocity distributions
 as well as
 the accumulated and the average kinetic energies of
 the (simulated) incident Galactic WIMPs
 {\em scattering} off target nuclei
 in the laboratory--independent Equatorial
 and Galactic coordinate systems,
 respectively.

 Five (spin--sensitive) nuclei
 used frequently
 in (directional) direct detection experiments:
 $\rmF$,
 $\rmAr$,
 $\rmGe$,
 $\rmXe$,
 and $\rmW$
 have been considered
 as our targets%
\footnote{
 Although Ar, Ge, and Xe are (so far) not used
 in directional detection experiments,
 for readers' cross reference to Ref.~\cite{DMDDD-NR},
 we present simulations with them here,
 which would show similar results
 as with the $\rmXA{S}{32}$/$\rmXA{Cl}{35/37}$,
 $\rmXA{Br}{79/81}$,
 and $\rmXA{I}{127}$ nuclei,
 respectively.
}.
 Then,
 while
 the SI (scalar) WIMP--nucleon cross section
 has been fixed as $\sigmapSI = 10^{-9}$ pb
 in our simulations
 presented in this paper,
 the effective SD (axial--vector) WIMP--proton/neutron couplings
 have been tuned as $\armp = 0.01$
 and $\armn = 0.7 \armp = 0.007$,
 respectively
 \cite{DMDDD-3D-WIMP-N}.
 So that
 the contributions of
 the SI and the SD WIMP--nucleus cross sections
 (including the corresponding nuclear form factors)
 to the validation criterion (\ref{eqn:f_NR_thetaNRchi})
 are approximately comparable
 for the considered $\rmAr$, $\rmGe$, and $\rmXe$ target nuclei%
\footnote{
 The mass of the $\rmF$ and the $\rmW$ nuclei
 are either too light or too heavy.
 Thus,
 with the same simulation setup,
 the SD or the SI WIMP--nucleus cross section dominates.
}.

 Moreover,
 in this paper
 we assume simply that
 the experimental threshold energies
 for all considered target nuclei
 are negligible,
 whereas
 the maximal experimental cut--off energy
 has been set as $\Qmax = 100$ keV.
 5,000 experiments with 500 accepted
 events on average
 (Poisson--distributed)
 in one observation period
 (365 days/year
  or 60 days/season)
 in one experiment
 for one laboratory/target nucleus
 have been simulated.

 For readers' reference,
 all simulation plots presented in this paper (and more)
 can be found ``in animation''
 on our online (interactive) demonstration webpage
 \cite{AMIDAS-2D-web}.

\subsection{Target dependence of
            the 3-D WIMP effective velocity distribution}
\label{sec:fv_eff-Eq-target}
 \def \WIMPmass    {0100}
 \def \ShortFrame  {Eq}
 \def \EventNumber {0500}
 \def \PlotNumber  {\PlotNumberAa}
 \InsertPlotfv
  {target&mchi=100}
  {The radial (far--left)
   and the angular (center--left) components of
   the WIMP effective velocity distribution of
   the scattering WIMPs
   as well as
   the angular distributions of
   the corresponding accumulated (center--right)
   and average (far--right) WIMP kinetic energy
   in the Equatorial coordinate system.
   (Note that
    the angular distributions are
    in unit of the all--sky average values
    and the scale of the color bar
    used in the far--right column is different
    from that in the two central columns.)
   Five target nuclei:
   $\rmF$  (a),
   $\rmAr$ (b),
   $\rmGe$ (c),
   $\rmXe$ (d),
   and $\rmW$  (e),
   have been presented.
   The mass of incident WIMPs
   has been set as $\mchi = 100$ GeV.
   500 accepted
   WIMP scattering events on average
   in one entire year
   have been simulated
   and binned into 15 (radial) bins
   as well as
   12 $\times$ 12 (angular) bins
   for the azimuthal angle and the elevation,
   respectively.
   The dark--green star indicates
   the theoretical main direction of incident WIMPs
   in the Equatorial coordinate system
   \cite{Bandyopadhyay10}:
   42.00$^{\circ}$S, 50.70$^{\circ}$W.
   See the text for further details.
   \vspace{-1.15 cm}%
   }
\begin{figure} [p!]
\begin{center}
 \begin{subfigure} [c] {12.5 cm}
  \includegraphics [width = 12.5 cm] {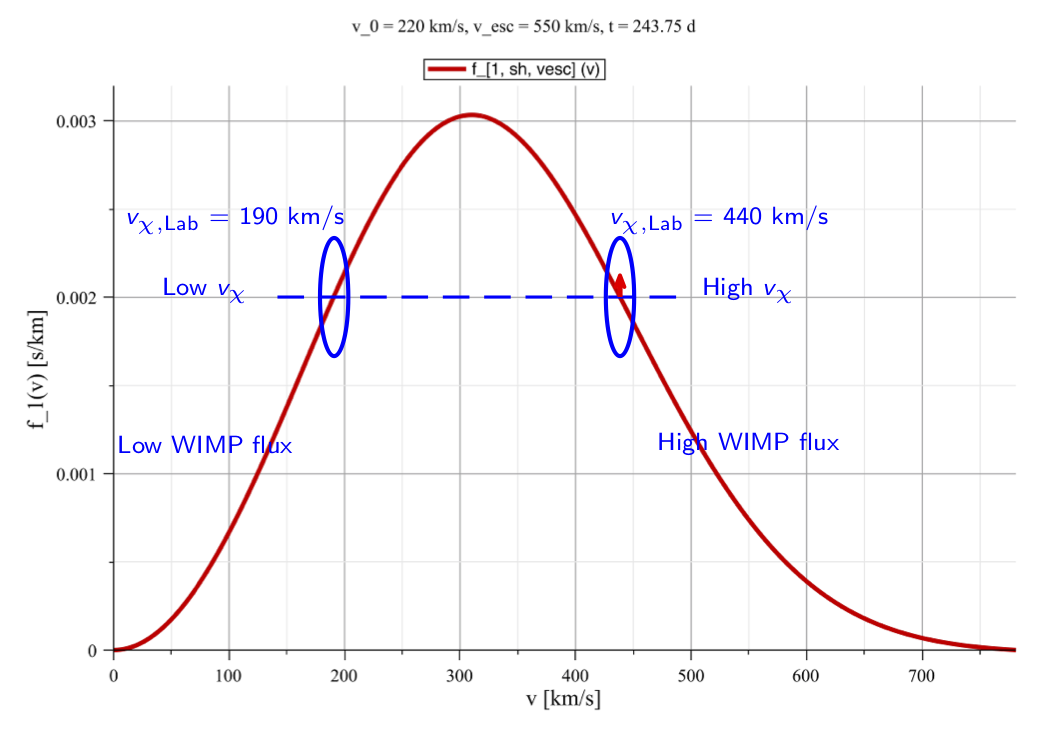}
 \caption{}
 \end{subfigure}
 \\
 \vspace{0.5 cm}
 \begin{subfigure} [c] {12.5 cm}
  \includegraphics [width = 12.5 cm] {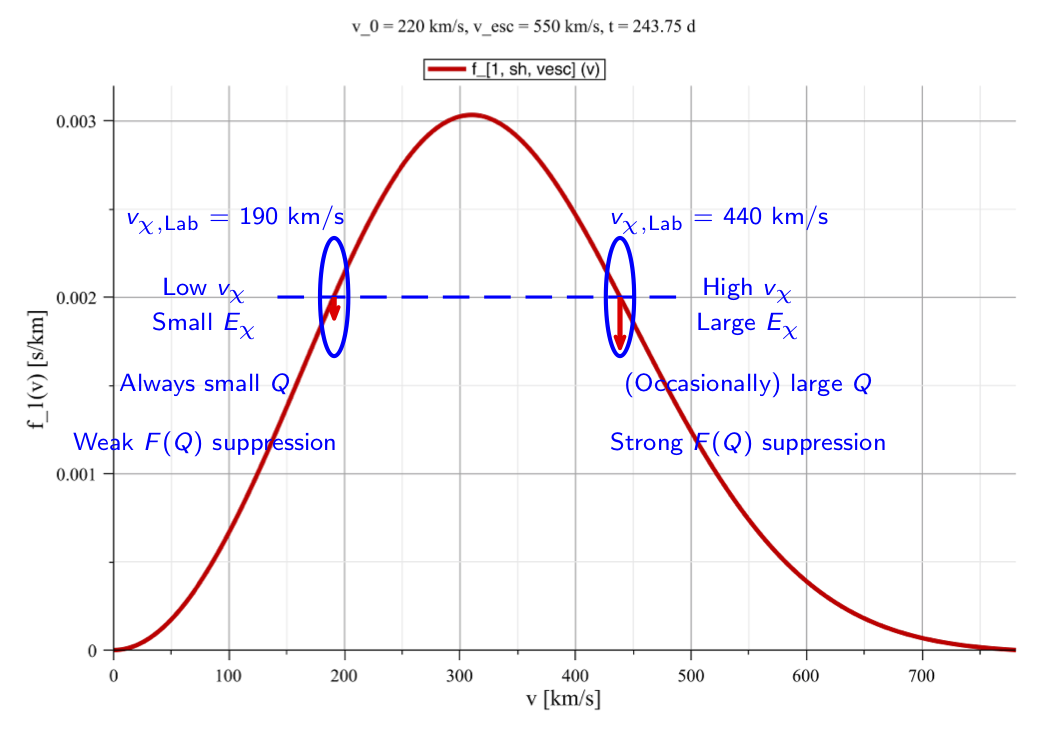}
 \caption{}
 \end{subfigure}
\end{center}
\caption{
 Two factors
 which cause
 the shift of
 the radial component (magnitude) of
 the 3-D WIMP effective velocity distribution:
 (a)
 the proportionality of the WIMP flux to the incident velocity
 (b)
 the cross section (nuclear form factor) suppression.
 See the text for the detailed arguments.
}
\label{fig:fv_eff-radial}
\end{figure}

 In Figs.~\ref{fig:fv-0100-Eq-0500-\PlotNumberAa},
 we show
 the radial (far--left)
 and the angular (center--left) components of
 the effective velocity distribution of
 the scattering WIMPs
 as well as
 the angular distributions of
 the corresponding accumulated (center--right)
 and average (far--right) WIMP kinetic energy
 in the Equatorial coordinate system
 (the angular distributions are
  in unit of the all--sky average values).
 Five target nuclei:
 $\rmF$,
 $\rmAr$,
 $\rmGe$,
 $\rmXe$,
 and $\rmW$,
 have been presented here%
\footnote{
 Interested readers can click each plot
 in Figs.~\ref{fig:fv-0100-Eq-0500-\PlotNumberAa}
 to open the corresponding webpage of
 the animated demonstration
 with varying target nuclei.
}
 and
 the mass of incident WIMPs
 has been set as $\mchi = 100$ GeV.
 500 accepted
 WIMP scattering events on average
 in one entire year
 have been simulated
 and binned into 15 (radial) bins
 as well as
 12 $\times$ 12 (angular) bins
 for the azimuthal angle and the elevation,
 respectively.
 The dark--green star in each plot indicates
 the theoretical main direction of incident WIMPs
 in the Equatorial coordinate system:
 \cite{Bandyopadhyay10}:
 42.00$^{\circ}$S, 50.70$^{\circ}$W.

 As a reference,
 in the far--left column
 of Figs.~\ref{fig:fv-0100-Eq-0500-\PlotNumberAa},
 we draw the theoretically predicted
 shifted Maxwellian velocity distribution of halo WIMPs
 as the solid red curves
 \cite{Lewin96}:
\beqn
 \conti
     f_{1, \sh, {\rm vesc}}(\vchiLab)
     \non\\
 \= {\footnotesize
     \cleft{\renewcommand{\arraystretch}{0.75}
            \begin{array}{l l}
             \D
             N_{\sh, {\rm vesc}}
             \afrac{\vchiLab}{v_0 \ve}
             \bBig{  e^{-(\vchiLab - \ve)^2 / v_0^2}
                   - e^{-(\vchiLab + \ve)^2 / v_0^2} } \~, &
             {\rm for}~\vchiLab \le \vesc - \ve        \~, \\
             ~ & ~ \\
             \D
             N_{\sh, {\rm vesc}}
             \afrac{\vchiLab}{v_0 \ve}
             \bBig{  e^{-(\vchiLab - \ve)^2 / v_0^2}
                   - e^{- \vesc^2           / v_0^2} } \~, &
             {\rm for}~\vesc - \ve \le \vchiLab \le \vesc + \ve \~, \\
             ~ & ~ \\
             0                                                  \~, &
             {\rm for}~\vchiLab \ge \vesc + \ve \~,
            \end{array}} }
\label{eqn:f1v_sh_vesc}
\eeqn
 with the normalization constant
\beq
     N_{\sh, {\rm vesc}}
  =  \bbrac{  \sqrt{\pi} \~
              \erf{\D \afrac{\vesc}{v_0}}
            - \afrac{2 \vesc}{v_0}
              e^{-\vesc^2 / v_0^2}        }^{-1}
\~.
\label{eqn:N_sh_vesc}
\eeq
 Here
 $\ve$ is the time--dependent
 Earth's velocity in the Galactic frame
 \cite{Freese88,
       SUSYDM96}:
\beq
     \ve(t)
  =  v_0 \bbrac{1.05 + 0.07 \cos\afrac{2 \pi (t - \tp)}{1~{\rm yr}}}
\~,
\label{eqn:ve}
\eeq
 with $\tp \simeq$ June 2nd,
 the date
 on which
 the Earth's orbital speed
 is maximal%
\footnote{
 As usual,
 the time dependence of $\ve(t)$
 has been ignored and
 $\ve = 1.05 \~ v_0$
 is used.
}$^{,\~}$%
\footnote{
 Note that,
 although
 it would be somehow inconsistent
 with our observations presented
 in Ref.~\cite{DMDDD-N}
 (see detailed discussions therein),
 we adopted the expression for $\ve(t)$
 as well as
 the date of $\tp$ here.
}.
 Additionally,
 the thin vertical dashed black lines
 here
 indicate the 1$\sigma$ Poisson statistical uncertainties
 on the recorded event numbers.

\begin{figure} [t!]
\begin{center}
 \includegraphics [width = 12 cm] {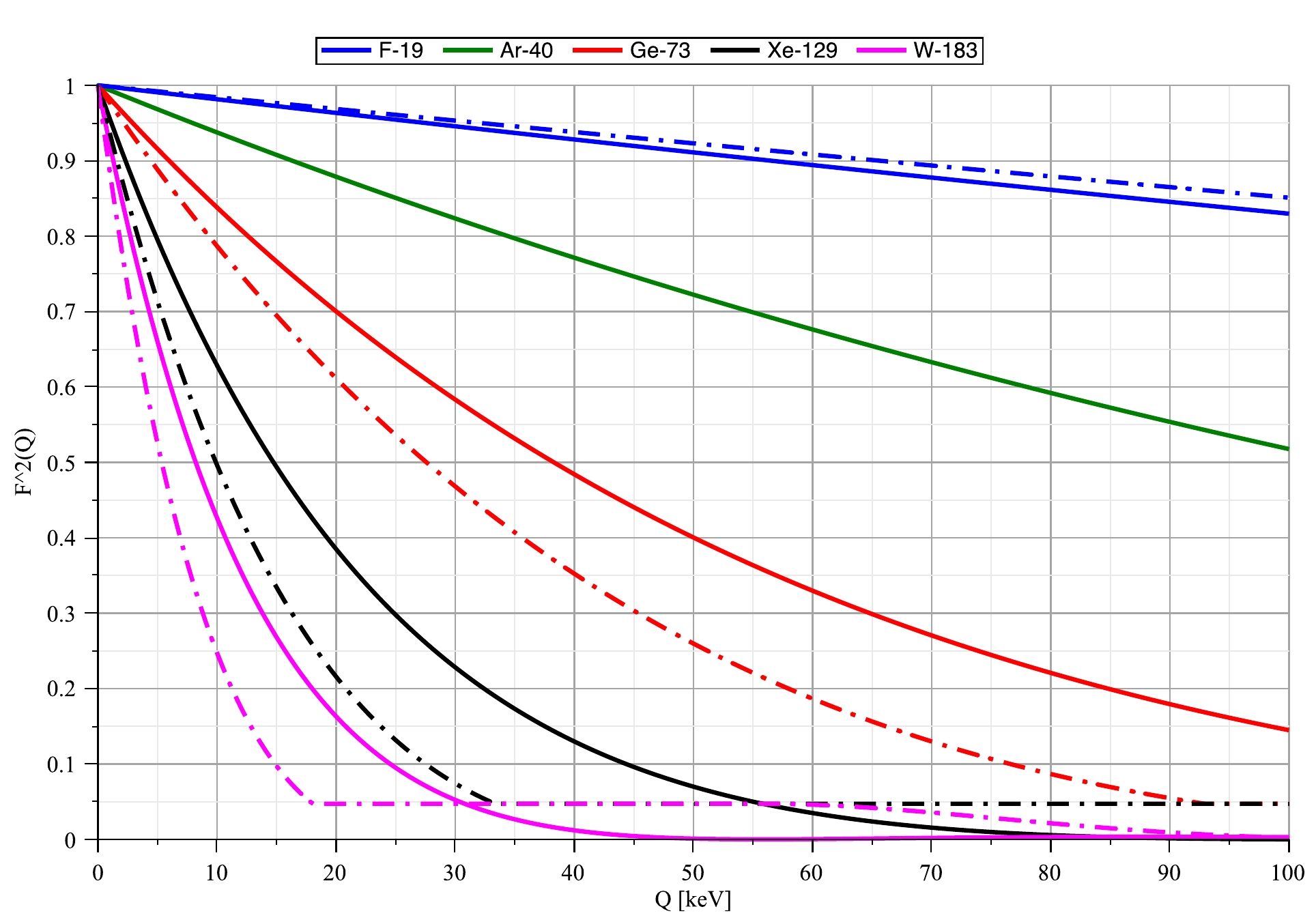}
\end{center}
\caption{
 Nuclear form factors of
 the $\rmF$     (blue),
 the $\rmAr$    (green),
 the $\rmGe$    (red),
 the $\rmXe$    (black),
 and the $\rmW$ (magenta) nuclei
 as functions of the recoil energy.
 The solid and dash--dotted curves
 indicate the form factors corresponding to
 the SI and SD cross sections
 adopted in our simulation package,
 respectively
 \cite{DMDDD-3D-WIMP-N}.
}
\label{fig:FQ}
\end{figure}

 Firstly,
 from the radial distribution (magnitude) of
 the incoming velocity of incident WIMPs
 scattering off target nuclei
 shown in Figs.~\ref{fig:fv-0100-Eq-0500-\PlotNumberAa},
 the target--dependent discrepancy between
 the simulated histogram
 (the {\em effective} WIMP velocity distribution)
 and the theoretically predicted
 (shifted Maxwellian) velocity distribution
 can be seen obviously.
 The reason is follows.

 Consider two subgroups of incident halo WIMPs
 with an equal velocity distribution value.
 For example,
 the moving velocities of two subgroup WIMPs are
 190 km/s
 and
 440 km/s,
 respectively.
 Then,
 as sketched in Fig.~\ref{fig:fv_eff-radial}(a),
 since
 the WIMP flux is proportional to the incident velocity,
 the higher the velocity,
 the larger the flux
 and in turn
 the larger the scattering probability
 off target nuclei.
 However,
 as also sketched in Fig.~\ref{fig:fv_eff-radial}(b),
 since
 the common velocity of
 the first subgroup of WIMPs
 is low
 and thus the common kinetic energy
 is small,
 they can only transfer small recoil energies
 to the scattered target nuclei.
 The reduction of
 the scattering probability of these WIMPs
 due to the (weak) cross section (nuclear form factor) suppression
 is then small or even negligible.
 In contrast,
 the common velocity of
 the second subgroup WIMPs
 is ($\sim$ 2.3 times) higher.
 So
 their common kinetic energy
 is ($\sim$ 5.4 times) larger
 and
 they could sometimes transfer (much) larger recoil energies
 to the scattered nuclei.
 However,
 the scattering probability of these large--recoil--energy events
 would be strongly reduced
 due to the sharply decreased cross section (nuclear form factor),
 especially
 with heavy target nuclei like $\rmXe$ and $\rmW$
 (see Fig.~\ref{fig:FQ}).

 This indicates that,
 due to the cross section (nuclear form factor) suppression
 appearing in Eq.~(\ref{eqn:f_NR_thetaNRchi}),
 the actual ``effective'' velocity distribution of
 incident halo WIMPs
 {\em scattering off} target nuclei
 should be (much) more strongly reduced
 from the theoretically predicted velocity distribution
 in the high velocity range
 than in the low velocity range.
 And
 the average velocity of
 the scattering WIMPs
 shifts to a lower velocity
 (than the theoretical prediction).
 In addition,
 as shown in Fig.~\ref{fig:FQ},
 the heavier the target nucleus,
 the stronger the cross section (nuclear form factor) suppression
 (in the high energy range)
 and thus
 the larger the shift of the average velocity.

 On the other hand,
 while
 the angular distribution patterns of
 the velocity direction (center--left)
 and the accumulated kinetic energy (center--right) of
 the scattering WIMPs
 look very similar to each other
 (but still slightly different)
 as well as
 to those of
 the entire incident WIMPs
 presented in Refs.~\cite{DMDDD-N, DMDDD-P},
 the differences between
 the distribution patterns of
 the average WIMP kinetic energy (far--right)
 off different target nuclei
 and that of
 the entire incident WIMPs
 \cite{DMDDD-P}
 seem somehow larger.
 Note here that,
 since
 the average velocity of
 WIMPs
 scattering off heavier target nuclei
 is lower
 and
 the all--sky average values of
 the accumulated and the average kinetic energies of
 the scattering WIMPs
 are in turn smaller,
 each distribution pattern
 shown in the same column
 of Figs.~\ref{fig:fv-0100-Eq-0500-\PlotNumberAa}
 is normalized by a different standard
 and demonstrates only the relative flux/kinetic energy of
 WIMPs with the simulated mass
 (e.g., $\mchi = 100$ GeV here)
 scattering off the considered target nucleus.

\subsection{WIMP--mass dependence of
            the 3-D WIMP effective velocity distribution}
\label{sec:fv_eff-Eq-mchi}
\begin{figure} [t!]
\begin{center}
 \includegraphics [width = 12 cm] {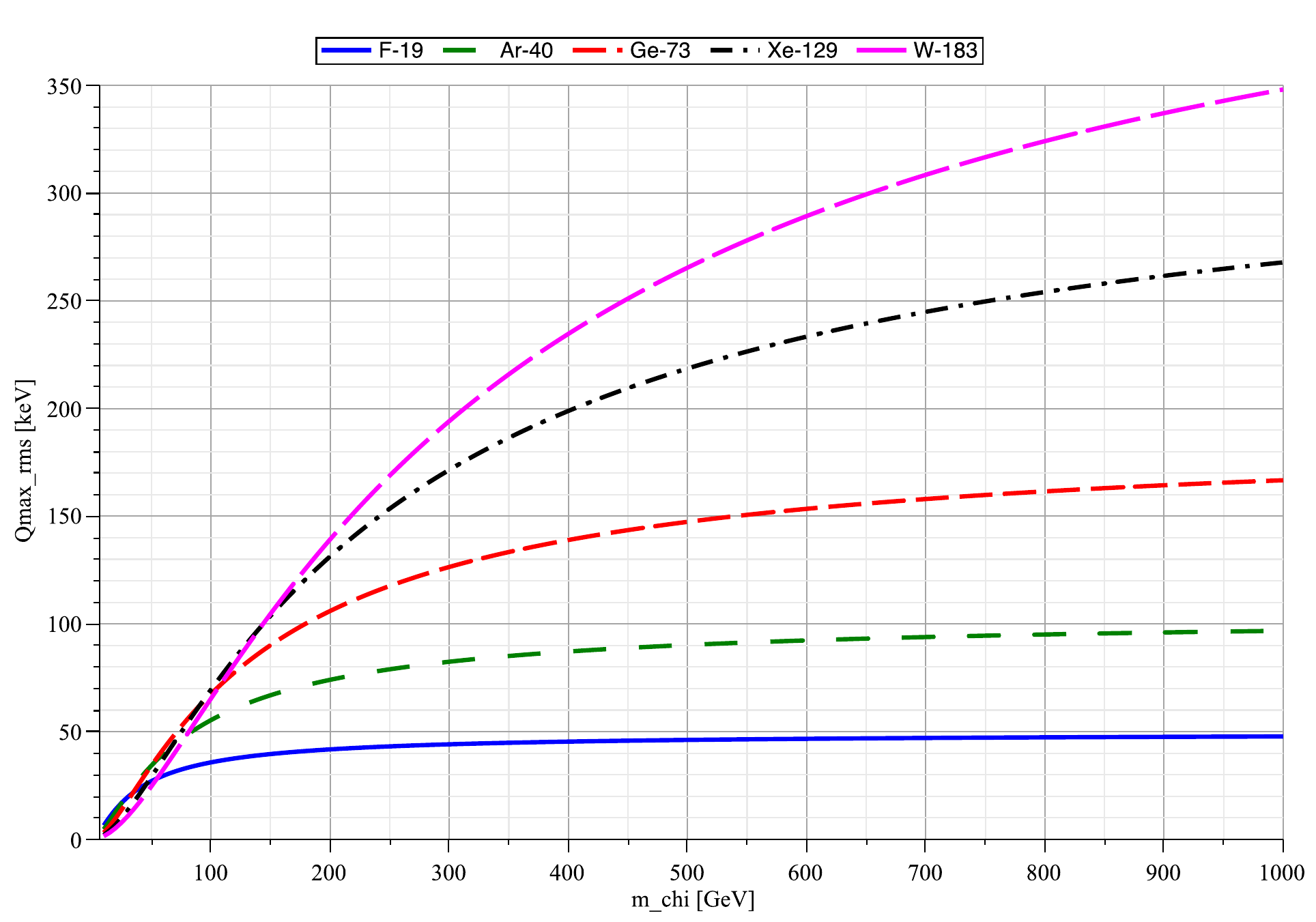}
\end{center}
\caption{
 The WIMP--mass dependence of
 the maximum of the recoil energy,
 $Q_{\rm max, rms}$,
 given by Eq.~(\ref{eqn:Qmax_rms}).
 Five frequently used target nuclei:
 $\rmF$     (solid        blue),
 $\rmAr$    (rare--dashed green),
 $\rmGe$    (dashed       red),
 $\rmXe$    (dash--dotted black),
 and $\rmW$ (long--dashed magenta)
 have been considered.
}
\label{fig:Qmax_rms-mchi}
\end{figure}
 \def \WIMPmass    {0020}
 \InsertPlotfv
  {mchi}
  {As in Figs.~\ref{fig:fv-0100-Eq-0500-\PlotNumberAa},
   except that
   the mass of incident WIMPs
   has been considered as light as $\mchi = 20$ GeV.%
   }
 \def \WIMPmass    {0200}
 \InsertPlotfv
  {mchi}
  {As in Figs.~\ref{fig:fv-0020-Eq-0500-\PlotNumberAa}
   and \ref{fig:fv-0100-Eq-0500-\PlotNumberAa},
   except that
   the mass of incident WIMPs
   has been considered as heavy as $\mchi = 200$ GeV.%
   }

 It is imaginable that,
 since
 the larger the mass of
 incident halo WIMPs,
 the larger
 their kinetic energy
 and
 the larger
 the maximal transferable recoil energy to
 the scattered target nuclei,
 the discrepancy
 between the effective velocity distribution of
 the scattering WIMPs
 and the theoretical prediction of
 the entire halo WIMPs
 caused by the cross section (nuclear form factor) suppression
 should be enlarged,
 once the WIMP mass becomes larger.
 More precisely,
 in Fig.~\ref{fig:Qmax_rms-mchi},
 the WIMP--mass dependence of
 the maximum (prefactor) of the recoil energy $Q$
 given by Eq.~(\ref{eqn:QQ_thetaNRchi})
 with the root--mean--square (rms) velocity
 \cite{DMDDD-3D-WIMP-N}
\beq
         v_{\rm rms, Lab}^2
  =      \afrac{3}{2} v_0^2 + \ve^2
 \simeq  (355~{\rm km/s})^2
\label{eqn:v2_rms_sh}
\eeq
 as the common velocity of incident WIMPs:
\beq
     Q_{\rm max, rms}
  =  \afrac{2 \mrN^2}{\mN} v_{\rm rms, Lab}^2
\~,
\label{eqn:Qmax_rms}
\eeq
 shows that
 the maximal recoil energy
 transferred by WIMPs
 increases basically with
 both of the increased mass of the target nucleus
 and that of incident WIMPs.
 Hence,
 in this subsection,
 we study
 the WIMP--mass dependence of
 the 3-D effective velocity distribution of
 the scattering WIMPs
 off different target nuclei
 in the Equatorial coordinated system
 in detail.

 In Figs.~\ref{fig:fv-0020-Eq-0500-\PlotNumberAa}
 and \ref{fig:fv-0200-Eq-0500-\PlotNumberAa},
 we show
 the radial (far--left)
 and the angular (center--left) components of
 the WIMP effective velocity distribution
 as well as
 the angular distributions of
 the accumulated (center--right)
 and average (far--right) WIMP kinetic energy
 (in unit of the all--sky average values)
 in the Equatorial coordinate system
 with
 a light WIMP mass of
 $\mchi = 20$ GeV
 and a heavy WIMP mass of 200 GeV,
 respectively%
\footnote{
 Interested readers can click each plot
 in Figs.~\ref{fig:fv-0020-Eq-0500-\PlotNumberAa}
 and \ref{fig:fv-0200-Eq-0500-\PlotNumberAa}
 to open the corresponding webpage of
 the animated demonstration
 with the varying WIMP mass.
}.

 It can be found that,
 once the mass of
 incident halo WIMPs
 is as light as ${\cal O}(20)$ GeV,
 the cross section (nuclear form factor) suppression
 is weak
 for all considered target nuclei
 and the factor of
 (the proportionality of the WIMP flux to)
 the incident velocity $\vchiLab$
 appearing in Eq.~(\ref{eqn:f_NR_thetaNRchi})
 dominates.
 Hence,
 the magnitudes of
 the ``20-GeV--WIMP'' effective velocity distribution
 off all considered target nuclei
 shift towards the high velocity range
 and the differences between them
 are pretty small.
 Raising the simulated WIMP mass,
 the peaks of
 the WIMP effective velocity distribution
 and the average/rms velocities of
 the scattering WIMPs
 shift to lower velocities.
 The heavier the mass of the target nucleus,
 the larger the shift
 and the lower the average/root--mean--square velocities.

 Meanwhile,
 the differences between
 the (characteristic) distribution patterns of
 the average WIMP kinetic energy (far--right)
 scattering off different target nuclei
 becomes also larger
 with the increased WIMP mass.
 For middle--mass and heavy target nuclei
 like $\rmGe$ and $\rmXe$,
 two extra hot--points
 close to the center and
 at the southwestern corner of the sky
 \cite{DMDDD-P}
 would be more obvious
 than for light target nuclei
 like $\rmF$ and $\rmAr$.%
\footnote{
 Remind here that,
 the all--sky average values of
 the accumulated/average kinetic energies of
 the scattering WIMPs
 are target (and WIMP--mass) dependent,
 and thus
 each distribution pattern
 shown in the same column
 of Figs.~\ref{fig:fv-0020-Eq-0500-\PlotNumberAa}
 and \ref{fig:fv-0200-Eq-0500-\PlotNumberAa}
 is normalized by a different standard.
}
\subsection{Annual modulation of
            the 3-D WIMP effective velocity distribution}
\label{sec:fv_eff-Eq-annual}

 Due to the orbital rotation of the Earth around the Sun,
 the relative velocity of incident halo WIMPs
 with respect to our laboratory/detector
 varies annually.
 Then
 the flux and the kinetic energy of incident WIMPs,
 the recoil energy of target nuclei
 scattered by incident WIMPs,
 as well as
 the WIMP effective velocity distribution
 should in turn vary annually.
 Hence,
 in this subsection,
 we discuss briefly
 the annual modulation of
 the 3-D effective velocity distribution of
 the scattering WIMPs
 in the Equatorial coordinated system.

 \def \Target           {F19}
 \def \WIMPmass         {0100}
 \def \Perioda          {\PeriodCa}
 \def \Periodb          {\PeriodCb}
 \def \Periodc          {\PeriodCc}
 \def \Periodd          {\PeriodCd}
 \def \PlotNumbera      {\PlotNumberCa}
 \def \PlotNumberb      {\PlotNumberCb}
 \def \PlotNumberc      {\PlotNumberCc}
 \def \PlotNumberd      {\PlotNumberCd}
 \InsertPlotfvAnnual
  {100}
  {As in Figs.~\ref{fig:fv-0100-Eq-0500-\PlotNumberAa}(a):
   $\rmF$ target nuclei
   scattered by 100-GeV WIMPs,
   except that
   500 accepted events on average
   in each 60-day observation period
   of four {\em advanced} seasons
   \cite{DMDDD-N, DMDDD-3D-WIMP-N}
   have been considered.
   Besides the dark--green star
   indicating
   the theoretical main direction of
   the WIMP wind,
   the blue--yellow point in each plot
   indicates
   the opposite direction of
   the Earth's movement in the Dark Matter halo
   on the central date of the observation period
   \cite{DMDDD-N}.%
   }
 \def \Target           {Ge73}
 \InsertPlotfvAnnual
  {100}
  {As in Figs.~\ref{fig:fv-F19-0100-Eq-0500-\PlotNumberCa},
   except that
   a middle--mass nucleus $\rmGe$
   has been considered as our target.%
   }
 \def \Target           {Xe129}
 \InsertPlotfvAnnual
  {100}
  {As in Figs.~\ref{fig:fv-F19-0100-Eq-0500-\PlotNumberCa}
   and \ref{fig:fv-Ge73-0100-Eq-0500-\PlotNumberCa},
   except that
   a heavy nucleus $\rmXe$
   has been considered as our target.%
   }

 In Figs.~\ref{fig:fv-F19-0100-Eq-0500-\PlotNumberCa},
 \ref{fig:fv-Ge73-0100-Eq-0500-\PlotNumberCa},
 and \ref{fig:fv-Xe129-0100-Eq-0500-\PlotNumberCa},
 we show
 the radial (far--left)
 and the angular (center--left) components of
 the WIMP effective velocity distribution
 as well as
 the angular distributions of
 the accumulated (center--right)
 and the average (far--right) WIMP kinetic energy
 (in unit of the all--sky average values)
 in the Equatorial coordinate system
 with
 500 accepted events on average
 in each 60-day observation period
 of four {\em advanced} seasons
 \cite{DMDDD-N, DMDDD-3D-WIMP-N}%
\footnote{
 Interested readers can click each row
 in Figs.~\ref{fig:fv-F19-0100-Eq-0500-\PlotNumberCa},
 \ref{fig:fv-Ge73-0100-Eq-0500-\PlotNumberCa},
 and \ref{fig:fv-Xe129-0100-Eq-0500-\PlotNumberCa}
 to open the webpage of
 the animated demonstration
 for the corresponding annual modulation
 (and for more considered WIMP masses
  and target nuclei).
}.
 100-GeV WIMPs
 have been simulated to scatter off
 $\rmF$,
 $\rmGe$,
 and $\rmXe$
 target nuclei,
 respectively.
 Besides the dark--green star
 indicating
 the theoretical main direction of
 the WIMP wind,
 the blue--yellow point in each plot
 indicates
 the opposite direction of
 the Earth's movement in the Dark Matter halo
 on the central date of the observation period
 \cite{DMDDD-N}.
 Additionally,
 for demonstrating
 the WIMP--mass dependence of
 the annual modulation of
 the 3-D WIMP effective velocity distribution
 in the Equatorial coordinate system,
 in Figs.~\ref{fig:fv-Xe129-0200-Eq-0500-\PlotNumberCa},
 we raised the mass of incident WIMPs
 (scattering off $\rmXe$)
 to \mbox{$\mchi = 200$ GeV}.
 Note that,
 the theoretically predicted
 shifted Maxwellian velocity distribution
 given in Eq.~(\ref{eqn:f1v_sh_vesc})
 with the time {\em independent} relation $\ve = 1.05 \~ v_0$
 has been used here
 for drawing the (fixed) solid red reference curves
 in the far--left column.

 \def \Target           {Xe129}
 \def \WIMPmass         {0200}
 \InsertPlotfvAnnual
  {200}
  {As in Figs.~\ref{fig:fv-Xe129-0100-Eq-0500-\PlotNumberCa}:
   $\rmXe$ has been considered as the target nucleus,
   except that
   the mass of incident WIMPs
   has been raised to $\mchi = 200$ GeV.%
   }

 It can be found that,
 firstly,
 for all considered target nuclei,
 the magnitude of
 the WIMP effective velocity distribution
 varies annually
 with maximal (minimal) average velocities
 in the advanced summer (winter).
 Meanwhile,
 the angular distributions of
 the velocity direction (flux)
 and
 the kinetic energy of
 the scattering WIMPs
 show the expected clockwise--rotated annual variations
 following the movement of
 the instantaneous theoretical main direction of incident WIMPs.
 From
 the (characteristic) distribution patterns of
 the average WIMP kinetic energy,
 one can even observe
 the target and WIMP--mass dependences.

%

%
 %
%
\section{3-D WIMP effective velocity distribution
         in the Galactic frame}
\label{sec:fv_eff-G}

 The goal of
 directional direct Dark Matter detection experiments
 should not be limited to
 an identification of
 positive (annual/diurnal modulated) anisotropy of
 WIMP--nucleus scattering events
 and discriminating them
 from theoretically (approximately) isotropic/incoming--direction--known
 background/astrophysical events.
 A more important goal
 would be to understand
 the astrophysical and particle properties of
 Galactic WIMPs
 as well as
 the structure of Dark Matter halo.
 Hence,
 as a start,
 we discuss
 the 3-D effective velocity distribution of
 incident halo WIMPs
 in the Galactic coordinate system
 in this section.

\subsection{Target dependence of
            the 3-D WIMP effective velocity distribution}
\label{sec:fv_eff-G-target}
\begin{figure} [t!]
\begin{center}
 \begin{subfigure} [c] {8.25 cm}
  \includegraphics [width = 8.25 cm] {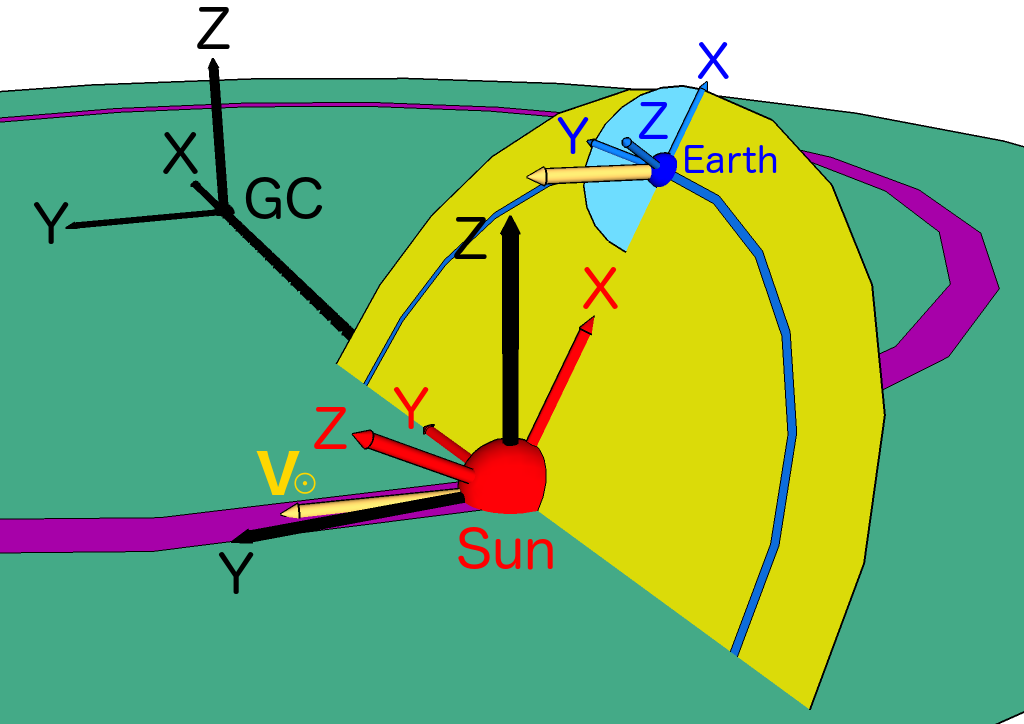}
 \caption{}
 \end{subfigure}
 \\
 \vspace{0.5 cm}
 \begin{subfigure} [c] {8.25 cm}
  \includegraphics [width = 8.25 cm] {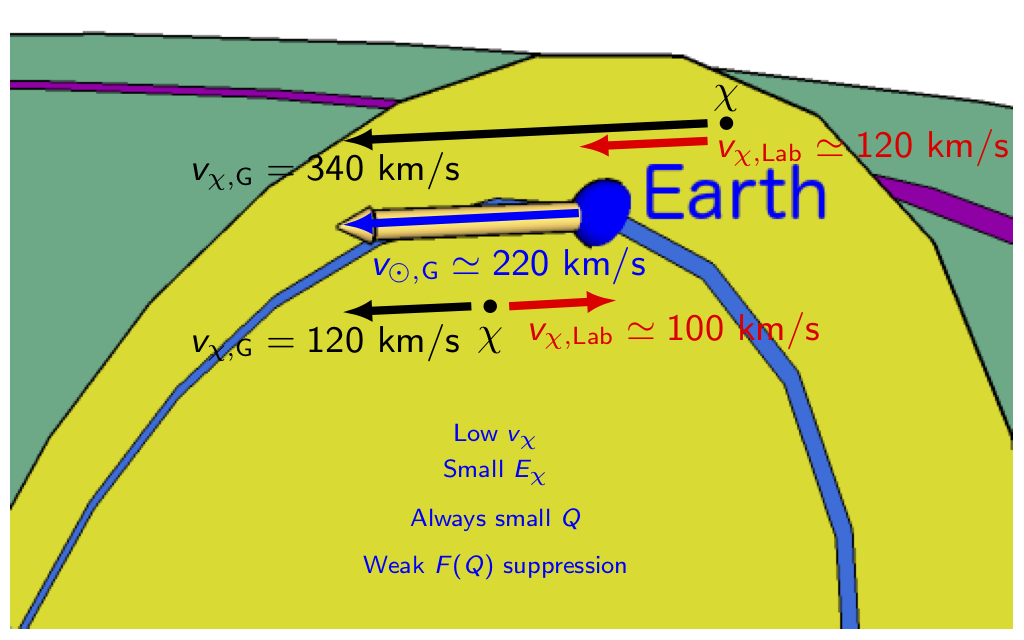}
 \caption{}
 \end{subfigure}
 \hspace{0.1 cm}
 \begin{subfigure} [c] {8.25 cm}
  \includegraphics [width = 8.25 cm] {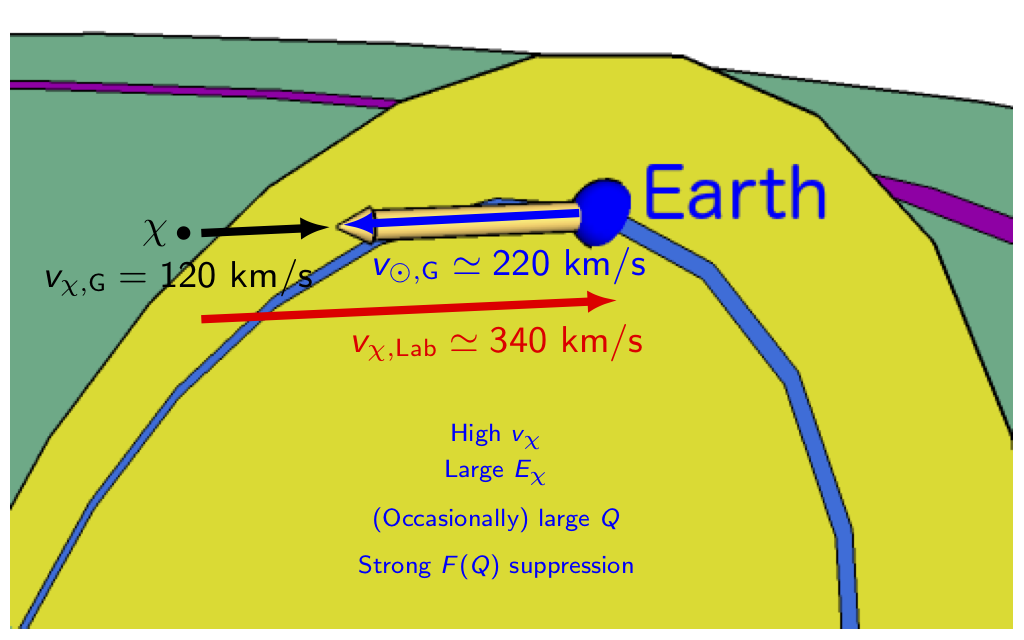}
 \caption{}
 \end{subfigure}
\end{center}
\caption{
 (a)
 The definitions of
 and the relative orientations between
 the (black) Galactic,
 the (red) Ecliptic,
 and the (blue) Equatorial coordinate systems
 (on the date of the vernal equinox)
 as well as
 the direction of the movement of the Solar system
 (the golden arrows)
 around the Galactic center
 (See Ref.~\cite{DMDDD-N} for details).
 The magenta circular band
 indicates an approximate path of
 the orbital motion of the Solar system
 in the Galaxy
 and
 the blue circular band indicates
 the Earth's orbit around the Sun.
 (b)(c)
 The explanations of
 the anisotropy (forward--backward asymmetry) of
 the incident flux/kinetic energy of
 the 3-D WIMP effective velocity distribution
 due to the cross section (nuclear form factor) suppression.
 See the text for the detailed arguments.
}
\label{fig:fv_eff-angular}
\end{figure}

 \def \WIMPmass    {0100}
 \def \ShortFrame  {G}
 \def \EventNumber {0500}
 \def \PlotNumber  {\PlotNumberAa}
 \InsertPlotfv
  {target&mchi=100}
  {As in Figs.~\ref{fig:fv-0100-Eq-0500-\PlotNumberAa}:
   100-GeV WIMPs scatter off
   $\rmF$  (a),
   $\rmAr$ (b),
   $\rmGe$ (c),
   $\rmXe$ (d),
   and $\rmW$ (e) nuclei,
   except that
   the 3-D information on
   the effective velocity distribution of
   the (simulated) incident halo WIMPs
   in the Galactic coordinate system
   has been presented.
   The solid red reference curves
   in the far--left column
   are the generating (simple Maxwellian) velocity distribution
   \cite{DMDDD-3D-WIMP-N}.
   And
   the dark--green/purple star
   on the left--hand (western) sky of each plot indicates
   the theoretical main direction of incident WIMPs
   in the Galactic coordinate system
   \cite{DMDDD-N, DMDDD-3D-WIMP-N}:
   0.60$^{\circ}$S, 98.78$^{\circ}$W,
   whereas
   the magenta/dark--green diamond
   on the right--hand (eastern) sky of each plot indicates additionally
   the moving direction of the Solar system
   in the Galactic coordinate system
   \cite{DMDDD-N, DMDDD-3D-WIMP-N}:
   0.60$^{\circ}$N, 81.22$^{\circ}$E
   (see also Fig.~\ref{fig:fv_eff-angular}(a)).
   \vspace{-0.1 cm}%
   }

 As in Sec.~\ref{sec:fv_eff-Eq-target},
 we show
 the 3-D effective velocity distributions
 as well as
 the angular distributions of
 the corresponding accumulated and average kinetic energies of
 incident halo WIMPs
 scattering off five considered target nuclei:
 $\rmF$,
 $\rmAr$,
 $\rmGe$,
 $\rmXe$,
 and $\rmW$,
 in the Galactic coordinate system
 in Figs.~\ref{fig:fv-0100-G-0500-\PlotNumberAa}.%
\footnote{
 Interested readers can click each plot
 to open the corresponding webpage of
 the animated demonstration
 with varying target nuclei.
}
 The mass of incident WIMPs
 has been set as $\mchi = 100$ GeV
 and,
 as a reference,
 the solid red curves indicating
 the generating
 (simple Maxwellian) velocity distribution
 \cite{DMDDD-3D-WIMP-N}
 have also been given
 in the far--left column.
 While
 the dark--green/purple star
 on the left--hand (western) sky of each plot indicates
 the theoretical main direction of incident WIMPs
 in the Galactic coordinate system
 \cite{DMDDD-N, DMDDD-3D-WIMP-N}:
 0.60$^{\circ}$S, 98.78$^{\circ}$W,
 the magenta/dark--green diamond
 on the right--hand (eastern) sky of each plot indicates additionally
 the moving direction of the Solar system
 in the Galactic coordinate system
 \cite{DMDDD-N, DMDDD-3D-WIMP-N}:
 0.60$^{\circ}$N, 81.22$^{\circ}$E
 (see Fig.~\ref{fig:fv_eff-angular}(a)
  for a 3-dimensional sketch
  in the Galactic point of view).

 Now,
 in the Galactic coordinate system,
 not only
 the target--dependent discrepancy of
 the magnitude of
 the 3-D WIMP effective velocity distribution
 from the generating
 velocity distribution,
 but also
 a target--dependent {\em anisotropy} of
 the moving direction (incident flux)
 as well as
 those of
 the angular distributions of
 the accumulate and the average kinetic energies
 can be observed clearly.

 More precisely,
 firstly,
 the one--dimensional WIMP effective velocity distributions
 shown in the far--left column of
 Figs.~\ref{fig:fv-0100-G-0500-\PlotNumberAa}
 indicate that,
 as in the Equatorial coordinate system,
 the average Galactic velocity of
 incident 100-GeV halo WIMPs
 scattering off
 light (heavy) target nuclei like
 $\rmF$ and $\rmAr$ ($\rmXe$ and $\rmW$)
 should be {\em higher} ({\em lower}) than that of
 the entire halo WIMPs.%

 Secondly and more importantly,
 the angular distribution patterns of
 the flux and the accumulated kinetic energy
 shown in two central columns
 of Figs.~\ref{fig:fv-0100-G-0500-\PlotNumberAa}
 indicate that
 WIMPs moving (approximately)
 in the {\em same} direction as
 the Galactic movement of our Solar system
 would have lower (higher) probabilities
 to scatter off
 light (heavy) target nuclei like
 $\rmF$ and $\rmAr$ ($\rmXe$ and $\rmW$)
 than WIMPs moving (approximately)
 in the {\em opposite} direction of
 the movement of our Solar system.
 However,
 interestingly,
 the angular distributions of
 the average WIMP kinetic energy
 shown in the far--right column
 of Figs.~\ref{fig:fv-0100-G-0500-\PlotNumberAa}
 indicate that,
 for all considered target nuclei,
 the forwardly--moving (and scattering) WIMPs
 would have {\em larger} average kinetic energies (velocities)
 than the backwardly--moving WIMPs.
 Moreover,
 all distribution patterns
 in Figs.~\ref{fig:fv-0100-G-0500-\PlotNumberAa}
 show that
 such a ``forward--backward asymmetry'' of
 the 3-D WIMP effective velocity distribution
 in the Galactic coordinate system
 would itself be {\em asymmetric}:
 the increases (decreases) of
 the incident flux/kinetic energy of
 the forwardly--moving WIMPs
 are obviously (much) stronger than
 those of
 the backwardly--moving WIMPs.

 The (asymmetric) forward--backward asymmetry of
 the 3-D WIMP Galactic effective velocity distribution
 is caused by
 the proportionality of the WIMP flux to the incident velocity
 as well as
 the cross section (nuclear form factor) suppression.
 Consider three cases.
 The first case:
 as sketched in Fig.~\ref{fig:fv_eff-angular}(b),
 our Solar system moves
 in the Galactic coordinate system
 with a velocity of $v_{\rm \odot, G} \simeq 220$ km/s
 and
 a group of WIMPs moves
 (approximately) in the same direction
 with a small common velocity of $\vchiG = 120$ km/s.
 While,
 in the Galactic point of view,
 our detector would catch up WIMPs in this group from behind,
 in the laboratory/detector points of view,
 these WIMPs would hit our detector head--on
 with a relative velocity of $\vchiLab \simeq 100$ km/s.
 On the other hand,
 in the second case,
 a group of WIMPs moves also
 in the (approximately) same direction
 as our Solar system
 but with a larger common velocity of $\vchiG = 340$ km/s.%
\footnote{
 For our generating simple Maxwellian velocity distribution
 \cite{DMDDD-3D-WIMP-N},
 one can find that
\beq
        f_{\chi, {\rm G, r}}(\vchiG = 120~{\rm km/s})
 \cong  f_{\chi, {\rm G, r}}(\vchiG = 340~{\rm km/s})
 \cong  0.0023~{\rm s/km}.
\eeq
}
 In both of the Galactic and the laboratory/detector points of view,
 these WIMPs would catch up and hit our detector from behind
 with a relative velocity of $\vchiLab \simeq 120$ km/s.
 In both cases,
 as argued in Sec.~\ref{sec:fv_eff-Eq-target},
 the relative velocities of
 incident WIMPs
 are low
 and
 their kinetic energies
 are small.
 Thus the WIMP fluxes are low,
 and the recoil energies
 transferred to the scattered target nuclei
 are always small,
 so that
 the cross section (nuclear form factor) suppression
 on the scattering probability of these WIMPs
 is weak or even negligible.

 In contrast,
 consider the third case that
 a group of WIMPs moves
 in the opposite direction
 towards our Solar system
 with a common velocity as low as $\vchiG = 120$ km/s
 (see Fig.~\ref{fig:fv_eff-angular}(c)).
 These WIMPs would hit our detector head--on
 with an ($\sim$ 3 times) higher incident velocity of
 $\vchiLab \simeq 340$ km/s
 and in turn
 a ($\sim$ 9 times) larger kinetic energy.
 So
 the WIMP fluxes are higher
 and the transferable recoil energies
 to the scattered target nuclei
 are (much) larger
 but practically almost always be suppressed,
 due to the (much) smaller
 cross section (nuclear form factor).

 Furthermore,
 the ``asymmetry'' of the forward--backward asymmetry of
 the WIMP Galactic effective velocity distribution
 is also easy to understand.
 The relative velocities of the backwardly--moving WIMPs
 with respective to our laboratory/detector
 are at least $\sim$ 220 km/s.
 In contrast,
 the relatively velocity of the forwardly--moving WIMPs
 can be very low
 and the varying ranges of
 their incident fluxes,
 kinetic energies,
 and transferable recoil energies
 are much larger. 

 It would be worth to mention here that,
 although the radial component of
 the 100-GeV WIMP effective velocity distribution
 off $\rmGe$ nuclei
 shown in Figs.~\ref{fig:fv-0100-G-0500-\PlotNumberAa}(c)
 seems to match the generating velocity distribution perfectly,
 the anisotropy of
 the angular distribution of
 the average kinetic energy
 could still be observed clearly.

\subsection{WIMP--mass dependence of
            the 3-D WIMP effective velocity distribution}
\label{sec:fv_eff-G-mchi}

 As in Sec.~\ref{sec:fv_eff-Eq-mchi},
 in this subsection,
 we consider
 the WIMP--mass dependence of
 the 3-D WIMP effective velocity distribution
 in the Galactic coordinated system.

 \def \WIMPmass    {0020}
 \InsertPlotfv
  {mchi}
  {As in Figs.~\ref{fig:fv-0100-G-0500-\PlotNumberAa},
   except that
   the mass of incident WIMPs
   has been considered as light as $\mchi = 20$ GeV.%
   }
 \def \WIMPmass    {0200}
 \InsertPlotfv
  {mchi}
  {As in Figs.~\ref{fig:fv-0020-G-0500-\PlotNumberAa},
   and \ref{fig:fv-0100-G-0500-\PlotNumberAa},
   except that
   the mass of incident WIMPs
   has been considered as heavy as $\mchi = 200$ GeV.%
   }

 In Figs.~\ref{fig:fv-0020-G-0500-\PlotNumberAa}
 and \ref{fig:fv-0200-G-0500-\PlotNumberAa},
 we show
 the 3-D effective velocity distributions
 as well as
 the angular distributions of
 the corresponding accumulated and average kinetic energies of
 incident halo WIMPs
 scattering off five considered target nuclei
 in the Galactic coordinate system.
 A light WIMP mass of
 $\mchi = 20$ GeV
 and a heavy one of 200 GeV
 have been simulated%
\footnote{
 Interested readers can click each plot
 in Figs.~\ref{fig:fv-0020-G-0500-\PlotNumberAa}
 and \ref{fig:fv-0200-G-0500-\PlotNumberAa}
 to open the corresponding webpage of
 the animated demonstration
 with the varying WIMP mass.
}.

 It can be seen clearly that,
 firstly,
 for the case of the light WIMP mass of $\mchi = 20$ GeV,
 the proportionality of the incident flux to the WIMP velocity dominates
 and thus
 the forwardly--moving WIMPs
 would have lower scattering probabilities
 off all considered target nuclei,
 but larger average kinetic energies,
 compared to the all--sky average values.
 Secondly,
 as already discussed in Sec.~\ref{sec:fv_eff-Eq-mchi},
 once the mass of incident WIMPs becomes larger,
 the differences between
 the one--dimensional
 WIMP effective velocity distribution
 as well as
 the angular distribution patterns of
 the velocity direction/kinetic energy of
 the scattering WIMPs
 off different target nuclei
 would indeed be more and more obvious.
 It would be worth to note here that,
 not like the distribution patterns
 in the Equatorial coordinate system
 demonstrated in Sec.~\ref{sec:fv_eff-Eq},
 even for the case of the light WIMP mass of $\mchi = 20$ GeV,
 the target dependence of
 the angular distribution patterns of
 the velocity direction/kinetic energy
 in the Galactic coordinate system
 could be observed clearly.

\subsection{Annual modulation of
            the 3-D WIMP effective velocity distribution}
\label{sec:fv_eff-G-annual}

 In Sec.~\ref{sec:fv_eff-Eq-annual},
 we have demonstrated
 the (target and WIMP--mass dependent) annual modulation of
 the 3-D WIMP effective velocity distribution
 off different target nuclei
 in the Equatorial coordinate system.
 Considering the more clear
 target and WIMP--mass dependences
 and the interesting forward--backward asymmetry of
 the 3-D WIMP effective velocity distribution
 in the Galactic coordinate system,
 it would then be reasonable
 to study and demonstrate here
 its annual modulation.

 \def \Target           {F19}
 \def \WIMPmass         {0100}
 \def \Perioda          {\PeriodCa}
 \def \Periodb          {\PeriodCb}
 \def \Periodc          {\PeriodCc}
 \def \Periodd          {\PeriodCd}
 \def \PlotNumbera      {\PlotNumberCa}
 \def \PlotNumberb      {\PlotNumberCb}
 \def \PlotNumberc      {\PlotNumberCc}
 \def \PlotNumberd      {\PlotNumberCd}
 \InsertPlotfvAnnual
  {100}
  {As in Figs.~\ref{fig:fv-0100-G-0500-\PlotNumberAa}(a):
   $\rmF$ target nuclei
   scattered by 100-GeV WIMPs,
   except that
   500 accepted events on average
   in each 60-day observation period
   of four {\em advanced} seasons
   \cite{DMDDD-N, DMDDD-3D-WIMP-N}
   have been considered.
   Besides the dark--green/purple star
   indicating
   the theoretical main direction of
   the WIMP wind
   and the magenta/dark--green diamond
   indicating
   the direction of the Solar Galactic movement,
   (the blue/red--yellow point and)
   the red/blue--yellow square in each plot
   indicate
   the (opposite) direction of
   the Earth's movement in the Dark Matter halo
   on the central date of the observation period
   \cite{DMDDD-N}.%
   }
 \def \Target           {Ge73}
 \InsertPlotfvAnnual
  {100}
  {As in Figs.~\ref{fig:fv-F19-0100-G-0500-\PlotNumberCa},
   except that
   a middle--mass nucleus $\rmGe$
   has been considered as our target.%
   }
 \def \Target           {Xe129}
 \InsertPlotfvAnnual
  {100}
  {As in Figs.~\ref{fig:fv-F19-0100-G-0500-\PlotNumberCa}
   and \ref{fig:fv-Ge73-0100-G-0500-\PlotNumberCa},
   except that
   a heavy nucleus $\rmXe$
   has been considered as our target.%
   }

 In Figs.~\ref{fig:fv-F19-0100-G-0500-\PlotNumberCa},
 \ref{fig:fv-Ge73-0100-G-0500-\PlotNumberCa},
 and \ref{fig:fv-Xe129-0100-G-0500-\PlotNumberCa},
 we show
 the 3-D effective velocity distributions
 as well as
 the angular distributions of
 the accumulated and the average kinetic energies of
 the scattering WIMPs
 in the Galactic coordinate system
 with
 500 accepted events on average
 in each 60-day observation period
 of four advanced seasons%
\footnote{
 Interested readers can click each row
 in Figs.~\ref{fig:fv-F19-0100-G-0500-\PlotNumberCa},
 \ref{fig:fv-Ge73-0100-G-0500-\PlotNumberCa},
 and \ref{fig:fv-Xe129-0100-G-0500-\PlotNumberCa}
 to open the webpage of
 the animated demonstration
 for the corresponding annual modulation
 (and for more considered WIMP masses
  and target nuclei).
}.
 100-GeV WIMPs
 have been simulated to scatter off
 $\rmF$,
 $\rmGe$,
 and $\rmXe$
 target nuclei,
 respectively.
 Besides the dark--green/purple star
 indicating
 the theoretical main direction of
 the WIMP wind
 and the magenta/dark--green diamond
 indicating
 the direction of the Solar Galactic movement,
 (the blue/red--yellow point and)
 the red/blue--yellow square in each plot
 indicate
 the (opposite) direction of
 the Earth's movement in the Dark Matter halo
 on the central date of the observation period
 \cite{DMDDD-N}.

 \def \Target           {Xe129}
 \def \WIMPmass         {0200}
 \InsertPlotfvAnnual
  {200}
  {As in Figs.~\ref{fig:fv-Xe129-0100-G-0500-\PlotNumberCa}:
   $\rmXe$ has been considered as the target nucleus,
   except that
   the mass of incident WIMPs
   has been raised to $\mchi = 200$ GeV.%
   }

 Firstly,
 in contrast to the variations of
 the one--dimensional (incident) velocity distribution of
 the scattering WIMPs
 in the Equatorial (laboratory) coordinate system
 shown in the far--left columns
 of Figs.~\ref{fig:fv-F19-0100-Eq-0500-\PlotNumberCa}
 to \ref{fig:fv-Xe129-0200-Eq-0500-\PlotNumberCa},
 the average Galactic velocity of
 light WIMPs
 scattering off light target nuclei
 like $\rmF$ and $\rmAr$
 would be minimal (maximal)
 in the advanced summer (winter).
 However,
 once the WIMP mass is
 as large as $\mchi \gsim \~ {\cal O}(200)$ GeV,
 the average WIMP Galactic velocity
 scattering off heavy target nuclei like $\rmXe$ and $\rmW$
 would reversely become maximal (minimal)
 in summer (winter).
 In Table \ref{tab:vchiG_max},
 we give
 the summary of
 the (advanced) season,
 in which
 the average Galactic velocity of the scattering WIMPs
 off the considered target nucleus
 would be maximal.

\begin{table} [t!]
\begin{center}
\renewcommand{\arraystretch}{1.5}
 \begin{tabular}{|| c || c | c | c | c | c ||}
\hline
\hline
 \multicolumn{6}{|| c ||}
  {Season of the maximal average Galactic velocity of the scattering WIMPs} \\
\hline
\hline
 \multirow{2}{*}{\makebox[3 cm][c]{Target nucleus}} &
 \multicolumn{5}{ c ||}{WIMP mass} \\
\cline{2-6}
                               &
 \makebox[2.25 cm][c]{ 20 GeV} &
 \makebox[2.25 cm][c]{ 50 GeV} &
 \makebox[2.25 cm][c]{100 GeV} &
 \makebox[2.25 cm][c]{200 GeV} &
 \makebox[2.25 cm][c]{500 GeV} \\
\hline
\hline
 $\rmF$  & \multicolumn{5}{ c ||}{}         \\
 $\rmAr$ & \multicolumn{4}{ c   }{Winter} & \\
\cline{1-1}
\cline{6-6}
 $\rmGe$ & \multicolumn{4}{ c  |}{}       & $^{~}$Winter$^\dagger$        \\
\cline{1-1}
\cline{4-6}
 $\rmXe$ & \multicolumn{2}{ c  |}{\multirow{2}{*}{}}                      &
                                  \multirow{2}{*}{$^{~}$Winter$^\dagger$} &
           \multicolumn{2}{ c ||}{\multirow{2}{*}{Summer}}                \\
 $\rmW$  & \multicolumn{2}{ c  |}{}       &
                                          &
           \multicolumn{2}{ c ||}{}       \\
\hline
\hline
\end{tabular}
\end{center}
\caption{
 The summary of
 the (advanced) season,
 in which
 the average Galactic velocity of the scattering WIMPs
 off the considered target nucleus
 would be maximal.
\\
 $^\dagger$:
 Determined by
 the one--dimensional
 WIMP effective velocity distribution
 shown in the far--left columns
 of Figs.~\ref{fig:fv-F19-0100-G-0500-\PlotNumberCa}
 to \ref{fig:fv-Xe129-0200-G-0500-\PlotNumberCa}.
}
\label{tab:vchiG_max}
\end{table}

 Secondly and very importantly,
 it could also be found that
 the angular distribution patterns of
 the velocity direction/kinetic energy of
 the scattering WIMPs
 on the eastern sky
 indeed rotate
 (approximately) {\em counterclockwise}
 following
 (the red/blue--yellow square indicating)
 the instantaneous Earth's movement in the Dark Matter halo,
 while
 the distribution patterns
 on the western sky
 rotate {\em synchronously}
 (and approximately) {\em clockwise}
 following
 (the blue/red--yellow point indicating)
 the instantaneous theoretical main direction of incident WIMPs.
 All (light and heavy) considered target nuclei
 would demonstrate this subtle annual variation of
 the forward--backward asymmetry.

 Finally,
 for the sake of completeness,
 in Figs.~\ref{fig:fv-Xe129-0200-G-0500-\PlotNumberCa}
 we raise the mass of incident WIMPs to $\mchi = 200$ GeV
 for demonstrating
 the WIMP--mass dependence of
 the annual modulation of
 the 3-D WIMP effective velocity distribution
 in the Galactic coordinate system.
 As expected,
 the annual modulations of
 the anisotropy of
 the angular distribution patterns of
 the velocity direction/kinetic energy of
 the scattering WIMPs
 becomes more obvious.

\section{Summary}

 In this paper,
 as the counterpart of
 our study on
 the angular distributions of
 the recoil direction (flux)
 and the recoil energy of
 the (Monte Carlo simulated) WIMP--scattered target nuclei
 \cite{DMDDD-NR},
 we investigated
 the corresponding 3-dimensional
 effective velocity distribution of
 WIMPs
 scattering off target nuclei
 in different celestial coordinate systems.

 Besides the proportionality of
 the incident flux to the WIMP velocity,
 we took into account
 the cross section (nuclear form factor) suppression
 on the transferred recoil energy
 in our double--Monte Carlo scattering--by--scattering simulation of
 the 3-dimensional elastic WIMP--nucleus scattering process
 and demonstrated
 the 3-D WIMP effective velocity distributions
 off several frequently used target nuclei
 in the Equatorial and the Galactic coordinate systems,
 which,
 instead of the theoretical predictions of
 the entire group of Galactic Dark Matter particles,
 describe the actual
 velocity distributions of WIMPs
 scattering off (specified) target nuclei
 in an underground detector.

 Our simulations showed that,
 firstly,
 in both of the Equatorial and the Galactic coordinate systems,
 there are clear discrepancies between
 the radial components (magnitudes) of
 the 3-D WIMP effective velocity distributions
 and the theoretically predicted and generating
 one--dimensional velocity distributions of
 the entire group of incident WIMPs.
 Such discrepancies depend on
 the target nucleus
 as well as
 on the mass of incident halo WIMPs:
 once the WIMP mass is as small as only $\cal O$(20) GeV,
 the proportionality of
 the WIMP flux
 to the incident velocity
 dominates
 and the average velocity/kinetic energy of
 the scattering WIMPs
 off all considered target nuclei
 would be larger than
 those of the entire halo WIMPs;
 with the increased WIMP mass,
 the average velocity of
 the scattering WIMPs
 shifts towards to lower velocities;
 the heavier the mass of the target nucleus,
 the larger the shift.

 Secondly and more importantly,
 the angular components (directions/fluxes) of
 the 3-D WIMP effective velocity distributions
 as well as
 the angular distributions of
 the accumulated kinetic energy of
 the scattering WIMPs
 in not only the Equatorial/laboratory
 but also the Galactic coordinate systems
 show clear target and WIMP--mass dependent
 anisotropies,
 which indicate
 the forward--backward asymmetry of
 the scattering rate of
 incident halo WIMPs:
 WIMPs moving in the same direction as
 the Galactic movement of the Solar system
 or, 
 more precisely,
 that of our laboratory/detector
 would have a higher (lower) probability
 to scatter off heavy (light) target nuclei
 than WIMPs moving in the opposite direction of
 the moving direction of the Solar system and our laboratory/detector.
 However,
 interestingly,
 for all considered target nuclei,
 the forwardly--moving (and scattering) WIMPs
 would have larger average velocities/kinetic energies
 than the backwardly--moving WIMPs.

 Moreover,
 the magnitudes of
 the 3-D WIMP effective velocity distributions
 in not only the Equatorial/laboratory
 but also the Galactic coordinate systems
 vary annually
 with the target and WIMP--mass dependences.
 Interestingly,
 while
 the average velocity of
 the scattering WIMPs
 in the Equatorial coordinate system
 would be maximal (minimal)
 in the advanced summer (winter),
 the average WIMP velocity
 in the Galactic coordinate system
 would reversely be minimal (maximal)
 in summer (winter),
 except of heavy ($\mchi \gsim \~ {\cal O}(200)$ GeV) WIMPs
 scattering off heavy target nuclei like $\rmXe$ and $\rmW$.

 In our simulations presented in this paper,
 500 accepted WIMP--scattering events
 on average
 in one observation period
 (365 days/year
  or 60 days/season)
 in one experiment
 for one laboratory/target nucleus
 have been simulated.
 Regarding the observation periods
 considered in our simulations
 presented in this paper,
 we used several approximations
 about the Earth's orbital motion
 in the Solar system.
 First,
 the Earth's orbit around the Sun is perfectly circular
 on the Ecliptic plane
 and the orbital speed is thus a constant.
 Second,
 the date of the vernal equinox is exactly fixed
 at the end of the May 20th (the 79th day) of
 a 365-day year
 and the few extra hours
 in an actual Solar year
 have been neglected.
 Nevertheless,
 considering the very low WIMP scattering event rate
 and thus maximal a few (tens) of total (combined) WIMP events
 observed in at least a few tens (or even hundreds) of days
 (an optimistic overall event rate of $\lsim \~ {\cal O}(1)$ event/day)
 for the first--phase analyses,
 these approximations should be acceptable.

 In summary,
 we finally achieved
 the full Monte Carlo scattering--by--scattering simulation of
 the 3-dimensional elastic WIMP--nucleus scattering process
 and can provide experimentally measurable (pseudo)data:
 the 3-dimensional recoil direction
 and the recoil energy of
 the WIMP--scattered target nuclei
 as well as
 the 3-D velocity/kinetic energy of
 the scattering WIMPs.
 Several important (but unexpected) characteristics
 have been observed.
 Hopefully,
 this (and more works fulfilled in the future)
 could help our colleagues
 to develop analysis methods
 for understanding
 the astrophysical and particle properties of
 Galactic WIMPs
 as well as
 the structure of Dark Matter halo
 by using directional direct detection data.

\subsubsection*{Acknowledgments}

 The author would like to thank
 the pleasant atmosphere of
 the W101 Ward and the Cancer Center of
 the Kaohsiung Veterans General Hospital,
 where part of this work was completed.
 This work
 was strongly encouraged by
 the ``{\it Researchers working on
 e.g.~exploring the Universe or landing on the Moon
 should not stay here but go abroad.}'' speech.

%
%
%
 %
%

%
%

%

%
%
%
\end{document}